\documentclass[twoside,onecolumn,tightenlines, showpacs,aps]{revtex4}
\usepackage{graphicx}
\newcommand{\be}{\begin{equation}}
\newcommand{\ee}{\end{equation}}
\begin{document}

\title{Complete Fusion of Weakly Bound Cluster-Type Nuclei at Near Barrier Energies}

\author{M S Hussein$^1$,P R S Gomes$^2$, J Lubian$^2$, R Linares$^2$, L F Canto$^3$}
\affiliation{$^1$ Instituto de Estudos Avan\c{c}ados and Instituto de F\'isica, Universidade de S\~ao Paulo, CP 66318, cep 05314-970, S\~ao Paulo, S.P., Brazil }
\affiliation{$^2$ Instituto de Fisica, Universidade Federal Fluminense, Av. Litoranea s/n, Niter�i, R.J. 24210-340, Brazil}
\affiliation{$^3$ Instituto de Fisica, Universidade Federal do Rio de Janeiro, CP 68528, Rio de Janeiro, Brazil}

\begin{abstract}
We consider the influence of breakup channels on the complete fusion of
weakly bound cluster-type systems in terms of dynamic polarization potentials. It is 
argued that the enhancement of the cross section at sub-barrier energies may be consistent with
recent experimental observations that nucleon transfer, often leading to breakup, is
dominant compared to direct breakup. The main trends of the experimental complete
fusion cross sections are analyzed in the framework of the Dynamic Polarization Potential approach. The qualitative conclusions are supported by CDCC calculations including a sequential
breakup channel, the one neutron stripping of $^7$Li followed by the breakup of $^6$Li.
\end{abstract}
\keywords{Fusion, Transfer, Breakup, Weakly bound projectiles}
\pacs{24.10Eq, 25.70.Bc, 25.60Gc }

\maketitle

The fusion and breakup of weakly bound, cluster-type, nuclei, both stable and radioactive,  has been a subject of great interest in the last years \cite{Canto06,Liang,Keeley}. Several systems have been studied, both theoretically and experimentally, including stable weakly bound projectiles ($^6$Li, $^7$Li and $^9$Be) and radioactive halo-type projectiles, like $^{6,8}$He, $^{7,11}$Be, $^8$B, $^{17}$F, on targets with masses ranging from $^7$Li to $^{238}$U. The basic question is whether the coupling to the breakup process enhances or hinders the fusion cross section. First, it should be stated what is considered as fusion cross section. Is it the complete fusion of the projectile with the target (CF) or the total fusion (TF), defined as the sum of the complete fusion and the incomplete fusion (ICF), the latter being the fusion of part of the projectile fragments after the breakup with the target? Most of the fusion data reported in the literature are for TF, since it is very difficult to separate experimentally CF from ICF. Accordingly one should address the question as to what reference calculatiom one would obtain  possible enhancement or suppression  Also, different breakup effects may occur, like static and dynamic effects. The first one is related with different barrier characteristics, when compared with those for similar tightly bound systems and the latter is related to the coupling of the breakup channel, which feeds continuum states, and other direct reactions. Therefore, if one compares data with theoretical predictions, the choice of the bare interacting potential plays a major role, and contradictory conclusions can be drawn with the same data set depending on the potential used \cite{GomesPLB11}.

The accepted picture of the fusion and breakup of a cluster-type nucleus on a certain target, is based on the
following decoposition of the different processes involved: 1) the complete capture of the whole projectile by the target (complete prompt fusion), 2) the breakup of the projectile followed by the sequential capture of both fragments ( complete sequential fusion), 3) the breakup of the projectile followed by the capture of one of the fragments, while the other fragment flys by (incomplete fusion), and 4) the breakup of the projectile with no capture of any of the fragments (non-capture breakup). The above processes compete and the affacts each others. In particular the study of the complete fusion, defined as the sum of 1) and 2), and the total fusion, defined as the sum of 1) , 2), and 3), are in general very much influenced by the coupling to the non-capture breakup process (also called "elastic" breakup). In this contribution we discuss the above processes and supply evidence of the occurence of yet another process, which we may call 5) transfer followed by breakup. This has been confirmed recently by the Canberra group  \cite{Luong}, and should be considered as another process to reckon with in discussing fusion of weakly bound stable or radioactive nuclei.

In the analysis of experimental data, the usual procedure has been to define the background or henchmark cross section, which describes the tunneling of the given system with due reference given to its general overall geometrical features, as well as its charge and mass. This is accomplished through what has been known as the Universal Fusion Function (UFF), introduced and discussed extensivly in \cite{Gomes05-017, Canto09}. This function hinges on an appropriate rescaling of the Wong formula for fusion \cite{Wong}, given by, $
\sigma_F = \frac{R_{B}^2 \hbar\omega}{2E}\ln[1 + \exp{\left(\frac{2\pi}{\hbar\omega}(E - V_B)\right)}]$.

The Wong cross section is useful in describing the fusion of strctureless nuclei, in the vicinity of the Coulomb barrier, and fails at energies well below the barrier owing to the parabolic (symmetrical) barrier used in its derivation. The actual Coulomb barrier is quite asymetrical. The limits of thw Wong formulas are well known, at above barrier-energies it reduces to the geomatrical formula, $\sigma_F = \pi R_{B}^2\left[1 -\frac{E}{V_B}\right]$,
and as expected no reference to the curvature of the barrier is maintained. At below-barrier energies, the Wong formuls reduces to the usaual exponential tunneling form,$
\sigma_F = \frac{R_{B}^2 \hbar\omega}{2E}\exp{\left(\frac{2\pi}{\hbar\omega}(E - V_B)\right)}$. Again we remind the reader that the above formula is not appropriate since the barrier is not an inverted parabola. An appropriate use of the correct classical action should be emplyed at these sub-barrier energies \cite{Canto06}.

The rescaling of the Wong formula is done by defining the variable $x = \frac{E - V_B}{\hbar\omega}$ and defining the UFF through $F_0(x) = \frac{2E}{R_{B}^2 \hbar\omega}\sigma_F$, thus, $F_0(x) = \ln[1 + \exp{2\pi x}]$. This function is universal in the sense that no specfic reference to the system is made. When comparing with the reduced data, the experimental F(x) must include couplings to all important bound channels. Deviations from UFF would be traced to the breakup and transfer couplings not included in the coupled-channels (CDCC) calculation. A large body of data have analysed using the above picture. Note that F(x) $\rightarrow 2\pi x$ even for small positive $x$.
In the following we consider the particular
data for collisions of $^{6,7}$Li projectiles incident upon a $^{209}$Bi
target, which have been measured with high precision \cite{Dasgupta02}.

Figure 1 shows the ratio between the complete fusion cross section of the $%
^{6}$Li + $^{209}$Bi reaction and that for $^{7}$Li + $^{209}$Bi, as a
function of the center of mass energy divided by the fusion barrier
energies, obtained from the measured fusion barrier distributions \cite{Dasgupta02}. Above the barrier, the
stronger the couplings that lead to prompt breakup, the larger is the
suppression. Below the barrier, the couplings give barrier weight at lower
energies. Because of the exponential dependence of tunneling probabilities
on the barrier energy, this outweighs the linear reduction in cross-section
due to prompt breakup. The behaviour seen by plotting the ratio of cross
sections for the two reactions (Fig. 1) is consistent with this picture.

Figure 2 shows the renormalized
complete fusion function for these two systems. The renormalized
fusion functions are obtained using the Sao Paulo potential \cite{Chamon}.
One can observe that the renormalized experimental complete fusion functions
are below the UFF (full curve) at energies above the barrier. The main features of
the data are summarized as: (\textit{i}) CF cross sections are suppressed by
about 30\% at energies above the barrier; (\textit{ii}) CF cross sections at
sub-barrier energies are enhanced by nearly one order of magnitude for $^6$%
Li; (\textit{iii}) The above two effects are more pronounced for $^6$Li than
for $^7$Li. The behavior below the barrier can be traced to the aforementioned
process 5) transfer followed by breakup. This can be seen in figure 3, where the real part of the dynamic polarization potential (DPP) is calculated for this process, and it shows a significant increase in attraction, resulting in a lower barrier.

Thus, the inclusion of the neutron pickup (transfer) followed by breakup can explain the below-barrier enhancement seen in the fusion of $^{6}$Li on a $^{209}$Bi target. The content of this contribution is a summary of a recent publication \cite{Gomes13}.

\begin{figure}[t]
\centering  \includegraphics[width=12cm]{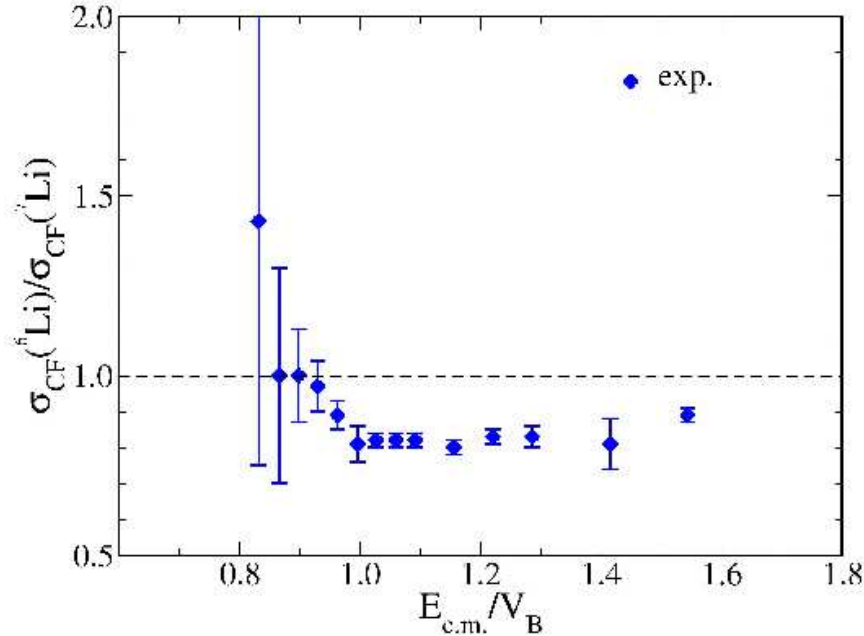} 
\caption{Ratio between the complete fusion cross section of the $^{6}$Li + $%
^{209}$Bi system and the one for the $^{7}$Li + $^{209}$Bi system, as a
function of the center of mass energy divided by the fusion barrier,
obtained from the measured fusion barrier distributions \protect\cite{Dasgupta02}.
Data from Ref.\protect\cite{Dasgupta02}}
\label{fig1}
\end{figure}

\begin{figure}[t]
\centering  \includegraphics[width=14cm]{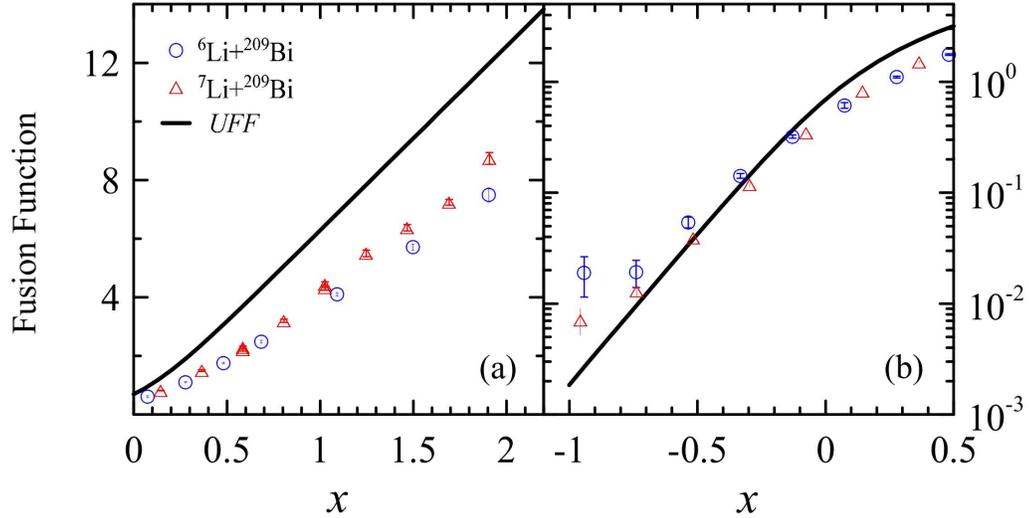} 
\caption{Renormalized fusion functions (see text) for complete fusion
plotted against $x$ = (E-V$_{B}$)/$\hbar \protect\omega $ for the two
systems. The data are from \protect\cite{Dasgupta02} and the full curves are the
universal fusion function (UFF) obtained by using the prescription of 
\protect\cite{Canto09}. }
\label{fig2}
\end{figure}

\begin{figure}[t]
\centering \includegraphics[width=10.5cm]{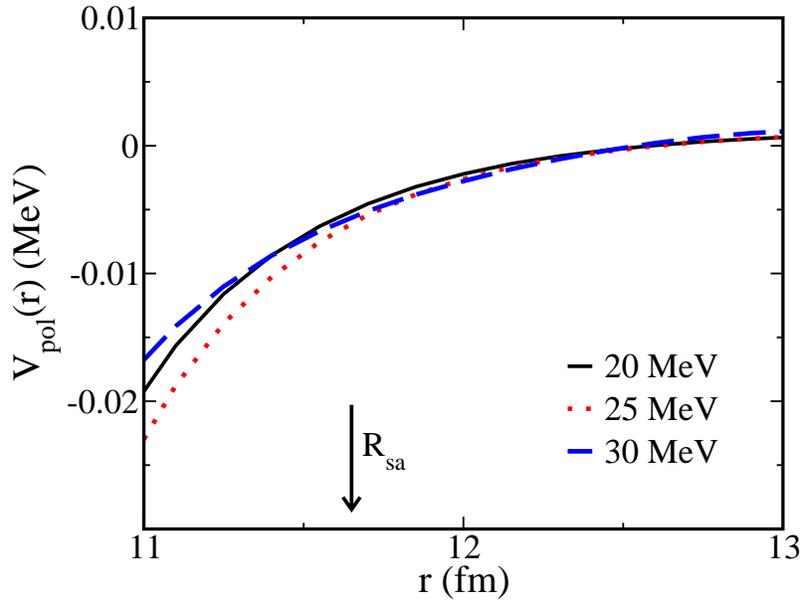} 
\caption{ Real part of the DPP around the strong
absorption radius for $^7$Li + $^{144}$Sm at laboratory energies above (30 MeV), below
(20 MeV) and close to the barrier (25 MeV) for breakup of $^6$Li into alpha + d following the
one-neutron stripping of $^7$Li.}
\label{fig3}
\end{figure}

\smallskip


\medskip
\begin{thebibliography}{9}
\bibitem{Canto06} Canto L F, Gomes P R S, Donangelo R, M. S. Hussein 2006 {\it Phys. Rep.} {\bf 424} 1
\bibitem{Liang} Liang J F, Signorini C 2005 {\it Int. J. Mod. Phys.} E {\bf 14} 1121
\bibitem{Keeley} Keeley N, Raabe R, Alamanos N, Sida J L 2007 {\it Prog. Part. Nucl. Sci.} {\bf 59} 579
\bibitem{GomesPLB11} Gomes P R S {\it et al.} 2011 {\it Phys. Lett.} B {\bf 695} 320
\bibitem{Gomes05-017} Gomes P R S, Lubian J, Padron I, Anjos R M 2005 {\it Phys. Rev.} C {\bf 71} 017601
\bibitem{Canto09} Canto L F {\it et al.} 2009 {\it J. Phys.} G {\bf 36} 015109;  2009 {\it Nucl. Phys} A {\bf 821}  51
\bibitem{Luong} Luong D H {\it et al.} 2011 {\it Phys. Lett.} B {\bf 695} 105 \textit{Europ. Phys. J. Web of Conf.} \textbf{17} 03002
\bibitem{Wong} Wong C Y 1973 {\it Phys. Rev. Lett.} {\bf 31} 766
\bibitem{Gomes09PRC} Gomes P R S, Lubian J, Canto L F 2009 {\it Phys. Rev.} C {\bf 79} 027606
\bibitem{Dasgupta02} Dasgupta M {\it et al} 2002 {\it Phys. Rev.} C {\bf 66} 041602(R) (2002); 2004 {\it Phys. Rev.} C {\bf 70} 024606
\bibitem{Chamon} Chamon L C {\it et al.} 2002 {\it Phys. Rev.} C {\bf 66} 014610; 1997 {\it Phys. Rev. Lett.} {\bf 79} 5218 
\bibitem{Gomes13} P. R. S. Gomes {\it et al.} 2013 {\it J. Phys. G: Nucl. Part. Phys.} {\bf 39} 115103
 

\end{thebibliography}
\end{document}